\documentclass[1p,numbers]{elsarticle}

\usepackage[english]{babel}
\usepackage[utf8x]{inputenc}
\usepackage{titlesec, amsmath,stackengine,graphicx,subfigure,tabularx,arydshln,url,stfloats}

\stackMath
\usepackage{multirow}
\usepackage[tableposition=top]{caption}
\usepackage[colorinlistoftodos]{todonotes}
\usepackage[colorlinks=true, allcolors=blue]{hyperref}
\usepackage{url}
\biboptions{sort&compress}
\title{Measurement of the onset of nucleate boiling in liquid xenon}
\begin{document}
\renewcommand\arraystretch{1.2}
\author[slac]{P.A.~Breur \corref{cor}}
\ead{sanderbreur@stanford.edu}
\author[pnnl,wsu]{G.S.~Ortega\corref{cor}}
\ead{gabriel.ortega@pnnl.gov}
\author[slac]{P.C.~Rowson}
\author[pnnl]{R.~Saldanha}
\author[slac]{B.~Mong}
\author[slac]{M.~Oriunno}
\author[wsu]{C.~Mo}

\cortext[cor]{Corresponding Authors}
\address[slac]{SLAC National Accelerator Laboratory, Menlo Park, California 94025, USA}
\address[pnnl]{Pacific Northwest National Laboratory, Richland, Washington 99354, USA}
\address[wsu]{Washington State University School of Mechanical and Materials Engineering, Richland, Washington 99354, USA}

\begin{abstract}
We report the first precision measurement of the superheat temperature required for bubble nucleation in liquid xenon of $\Delta T_{wall,ONB}$ = (16.9$\pm$0.5) K and $\Delta T_{wall,ONB}$~=~(19.2$^{+0.4}_{-1.1}$) K at pressures of P = $(0.98\pm0.02)$\,bar and P = $(1.32^{+0.05}_{-0.01})$\,bar, respectively. Both are measured at a subcooled bulk fluid temperature of $\sim$ 162\,K. A dedicated liquid xenon setup is used to measure the temperature at which bubble nucleation first appears on the flat surface of a resistor. The associated average heat flux at the surface of the resistor is determined using fluid dynamics simulations. Future experiments can use these experimental results to determine the likelihood of boiling in liquid xenon and to design effective thermal sinks that prevent bubble nucleation.

\end{abstract}
\maketitle
\section{Introduction}
The use of cryogenic fluids, such as liquid xenon (LXe), as a detection material in particle-physics experiments has rapidly increased over the past decade in the search for physics beyond the Standard Model. Several properties of xenon make it an optimal medium for rare event searches \cite{Auger:2012gs, Abe:2013tc,Aprile:2017aty,Akerib:2012ys,Cao:2014jsa,Gando:2012zm,Gomez-Cadenas:2013lta}. LXe is a transparent and dense fluid with a boiling temperature of around 165 K at atmospheric pressure. Due to its high density ($\rho = 2.8$\,g/cc) and high atomic number (Z=56) it is a detection medium which has high self-shielding properties against external $\gamma$–ray backgrounds. Both charged and neutral particles can create tracks of ionized and excited xenon atoms and dimers, leading to detectable free electrons and photons from which energy and location of the interaction are determined. As a scalable detector material, xenon has made it possible for the quick increase of total detector target mass from a few kilograms to ton scales over the past decade. 
One challenge with ton-scale detectors in general is that the large size leads to extended cable lengths between detector and readout electronics. This in turn leads to amplified electronic noise levels due to increased capacitance, and an increasing complexity of feedthrough design due to the larger number of readout channels. To address this problem, experiments such as nEXO \cite{Kharusi:2018eqi} aim to place the readout electronics within the LXe medium itself. A benefit to this approach is that the electronics are cooled by the cryogenic fluid. However, the heat dissipation from these electronics can create local boiling conditions due to surface temperatures being elevated above the liquid saturation temperature. The bubbles generated from boiling could cause microphonic noise in the electronics or cabling and lead to electrical breakdown in high electric fields. 

We perform dedicated measurements of the onset of nucleate boiling in LXe using a controlled heat source consisting of a thin film resistor with internal thermocouples immersed in LXe. Single-phase computational fluid dynamics (CFD) simulations are performed to determine the local increase of the temperature at the surface of the heat source exposed to LXe. The simulations expand on the notion that boiling starts at the saturation temperature for cryogenic liquids. This research shows that the localized formation of bubbles is reliant on raised surface temperatures above the saturation temperature.

\section{Pool Boiling in Cryogenic Fluids}
\begin{figure}[ht]
	\centering
	\captionsetup{}
	\includegraphics[width=\textwidth]{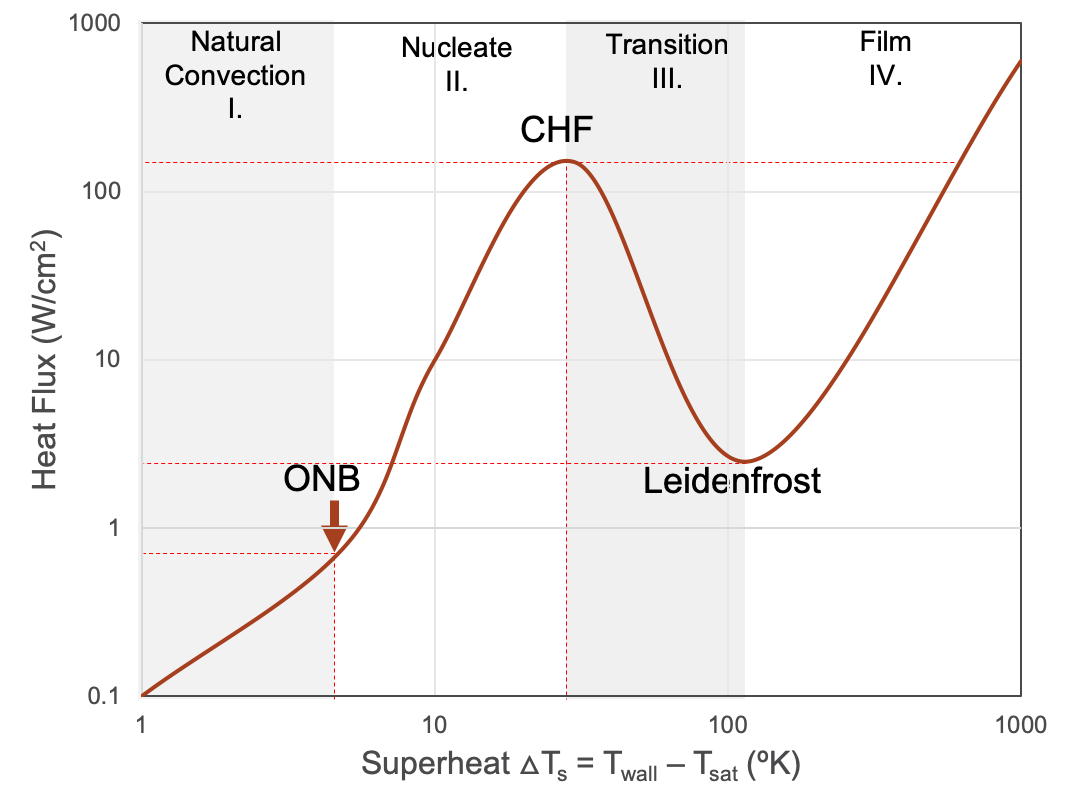}
	\caption{Example of the pool boiling curve for water showing different boiling regimes observed, data from \cite{cengel_heat_2019}. I) natural convection - heat transfer through fluid with vapor release at liquid-gas interface II) nucleate boiling - discrete bubble formation and detachment from surface III) transitional boiling - regions of the bulk fluid form larger bubbles IV) film boiling - vapor formation covers entire boiling surface creating an insulating layer. The Leidenfrost point signifies the transition to stable film boiling.}
	\label{fig:Pool Boiling Curve}
\end{figure}

 The term `pool boiling' refers to the initial fluid conditions being stagnant, as opposed to `flow boiling' where fluid is forced over the boiling surface. Figure \ref{fig:Pool Boiling Curve} depicts, as an example, a general boiling curve for water which shows the correlation of surface temperature ($T_{wall}$, the difference between surface and saturation temperature) to heat flux ($q$) for the four boiling regimes: natural convection, nucleate boiling, transitional boiling, and film boiling. Most studies in cryogenic liquids, like those done by Kida et al. \cite{kida_pool-boiling_1981} and Kosky and Lyon \cite{kosky_pool_1968} study the range from nucleate boiling (regime II in figure \ref{fig:Pool Boiling Curve}) up to the critical heat flux (CHF) where the heat transfer coefficient peaks. In this research we focus only on the detection of the onset of nucleate boiling (ONB) in LXe, defined by an increase of the heat transfer coefficient  when transitioning from a single phase to a two-phase flow regime (regime I in figure \ref{fig:Pool Boiling Curve}). 
In this work, the $q$ values and $T_{wall}$ above the saturation value are measured for a horizontal flat copper surface starting from a subcooled regime (initial fluid temperature below saturation temperature) at different pressures.

While the heat transfer coefficient, defined as $q/\Delta T$, for a single phase fluid under natural convection can be estimated by using the Nusselt number and thermal conductivity of the fluid, the determination of the super heating (equation \ref{eqn:delta_T_ONB}) at ONB ($\Delta T_{wall,ONB}$) requires experimental inputs.  The accepted phenomenological model of  nucleation boiling describes the superheating as the temperature in excess of saturation allowing  the local vapor pressure at the nucleation sites to overcome the surface tension \cite{warrier_onset_2002}. The onset of boiling is highly dependent on parameters such as the surface orientation, the material roughness and the amount of subcooling \cite{lamarsh_introduction_1983,bombardieri_influence_2016}. In water this value is generally accepted to be less than $5\,^{\circ}$K.  In LXe, the study by Haruyama \cite{haruyama_boiling_2002} finds $\Delta T_{wall,ONB}$ to be between approximately 3.8 and $18\,^{\circ}$K on a thin platinum wire, depending on hysteresis effects. This study aims to make a precise measurement of $\Delta T_{wall,ONB}$ in LXe on a flat copper surface.

\begin{equation}
\setstackgap{S}{2pt}
\Delta T_{wall,ONB} = T_{wall} - T_{sat}
\label{eqn:delta_T_ONB}
\end{equation}
 
\section{Experimental setup}
\begin{figure}[htb]
	\centering
	\captionsetup{}
	\includegraphics[width=0.7\textwidth]{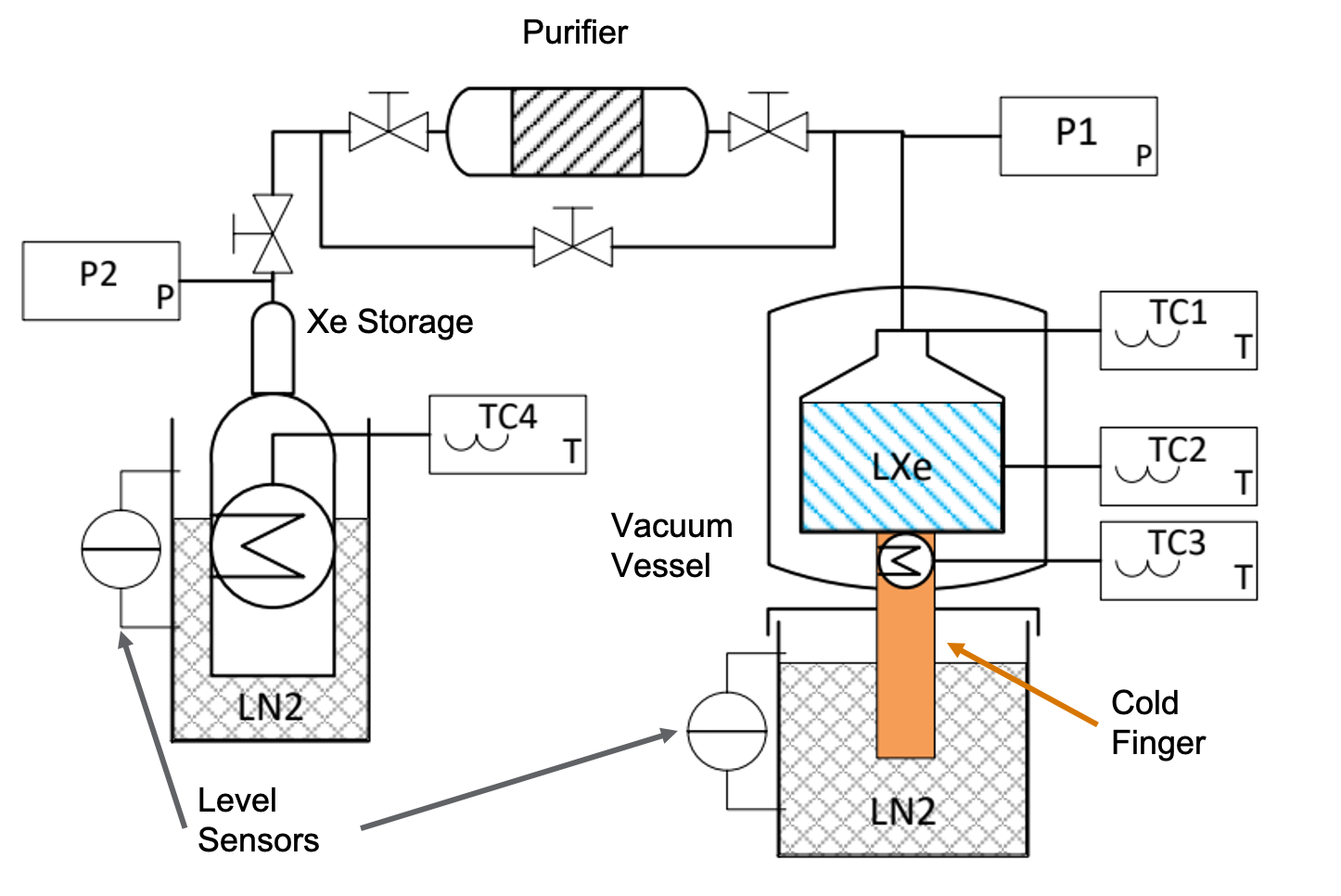}
	\caption{The simplified P\&I diagram of the experimental setup. From left to right: xenon storage bottle (with heater) within open liquid nitrogen dewar with temperature and pressure readout and control, purifier and bypass, liquid xenon test cell inside a vacuum chamber with thermal connection through cold finger (with heater) into liquid nitrogen. Several thermocouples and one pressure sensor are used to measure and control the liquid xenon test cell. Both liquid nitrogen dewars can be automatically refilled and are equipped with level sensors.}
	\label{fig:pid}
\end{figure}

\begin{figure}[htb]
	\centering
	\captionsetup{}
	\includegraphics[width=0.8\textwidth]{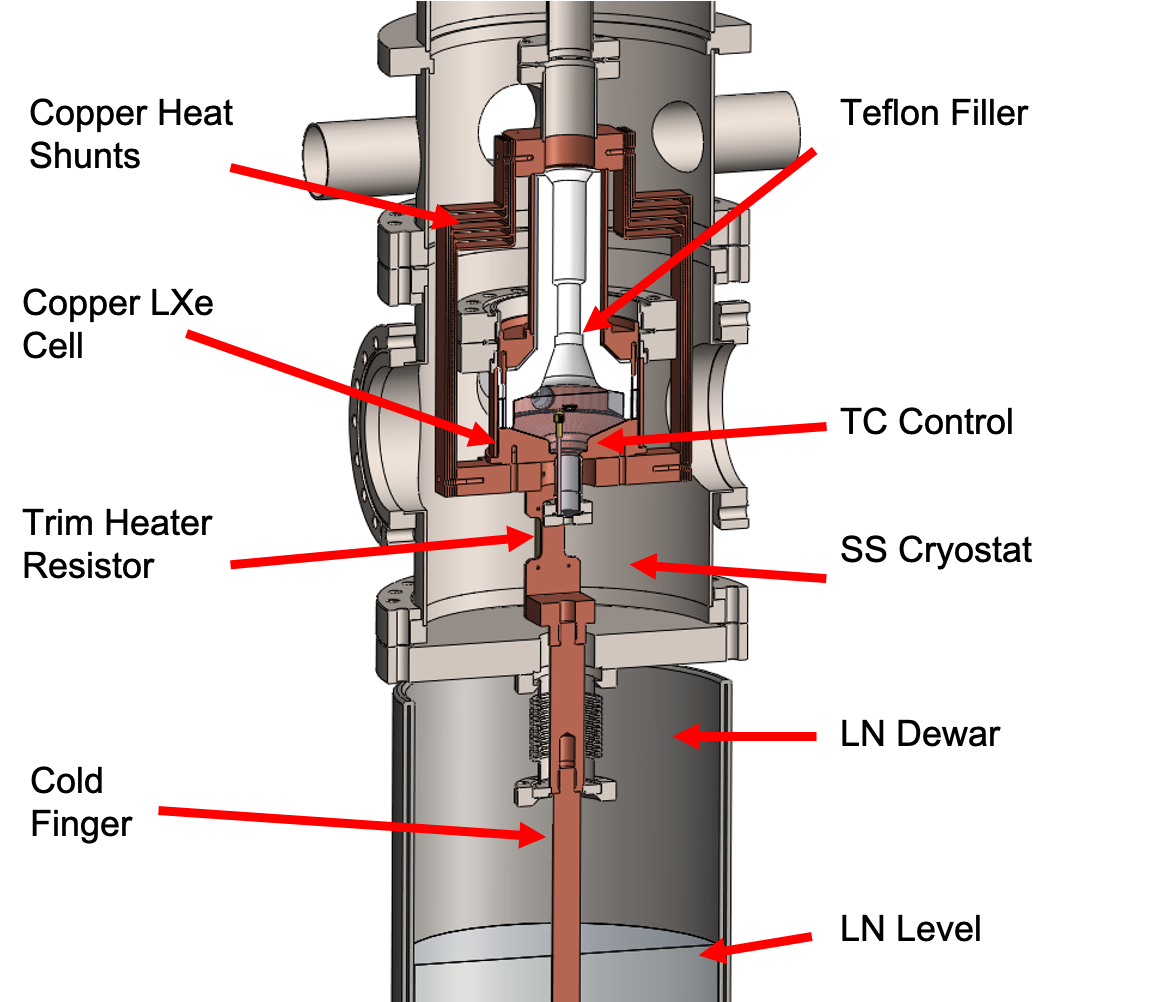}
	\caption{The LXe test cell cryostat, with the cold finger immersed in LN, the thermally-isolating cold finger feedthrough, the trim heater resistor (one of two – see Figure~\ref{fig:chamber_zoom}) indicated, and the inner copper LXe cell (shown in detail in Figure~\ref{fig:chamber_zoom}). Also visible are the copper heat shunts used to cool the upper portion of the cell, and the Teflon volume-displacing insert within the cell.}
	\label{fig:chamber_overview}
\end{figure}

\begin{figure}[ht]
	\centering
	\captionsetup{}
	\includegraphics[width=0.9\textwidth]{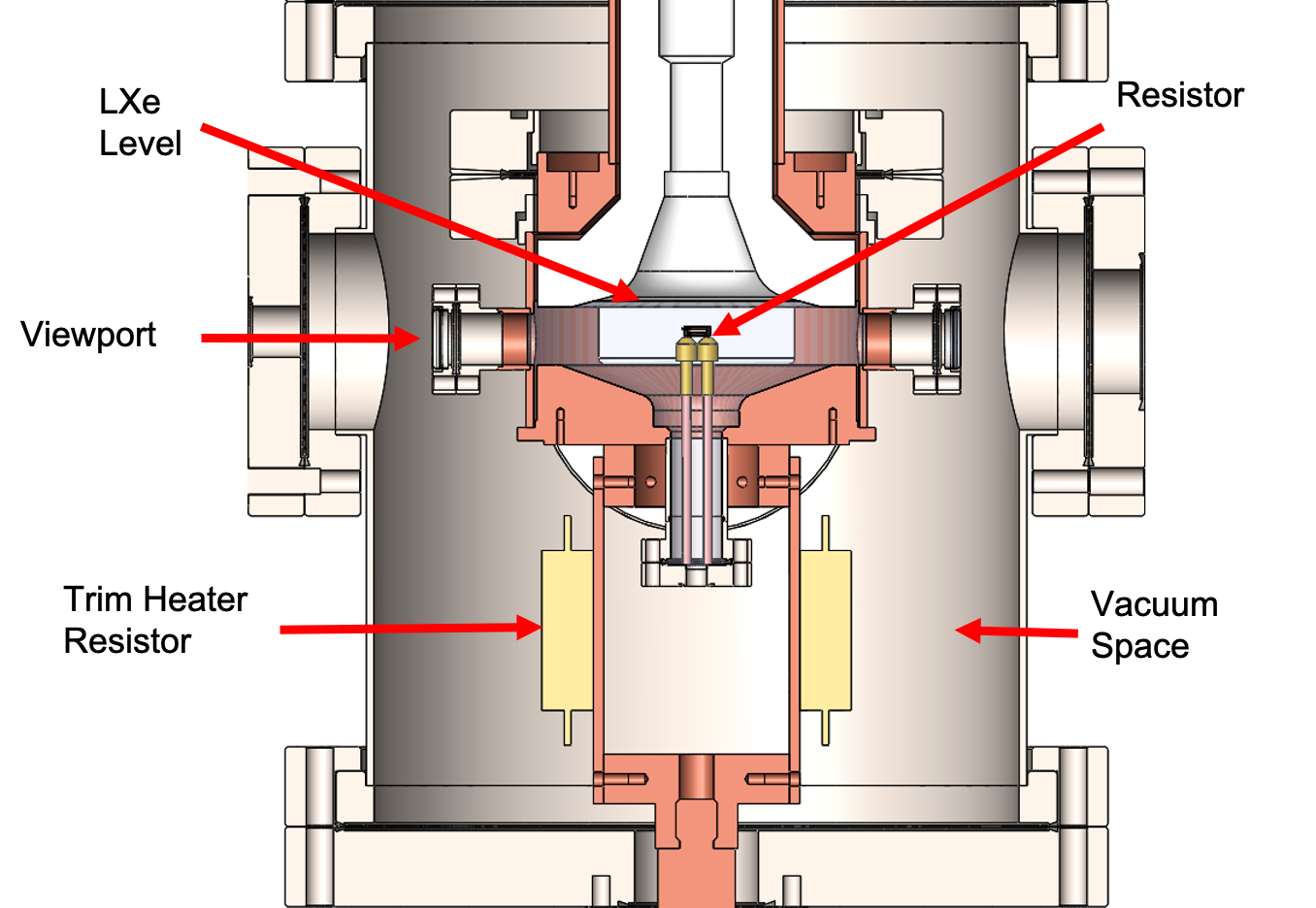}
	\caption{A close up of the test cell interior, showing the test resistor mounted to the  feedthrough, and the typical LXe level within the cell.  Bubble formation on the test resistor is viewed through the viewports and video clips were collected with a camera.}
	\label{fig:chamber_zoom}
\end{figure}

The experiments described here were performed in a LXe test cell at SLAC National Accelerator Laboratory. The setup has been used for a number of years \cite{rebel_high_2014} and was modified for these investigations. A simplified plumbing and instrumentation (P\&I) diagram is shown in figure \ref{fig:pid}. An open dewar is filled with liquid nitrogen (LN) whose level is maintained by a level sensor connected to a solenoid valve controller \cite{noauthor_custom_nodate} that automates refilling from a LN storage dewar (not shown). A cold finger immersed in LN is connected via a feedthrough into the insulating vacuum space to the LXe vessel, where a copper assembly is equipped with a pair of power resistor trim heaters (figures \ref{fig:chamber_overview} and \ref{fig:chamber_zoom}). 

Figure \ref{fig:chamber_overview} shows the overall layout of the cryostat and the test cell. The copper LXe cell assembly is instrumented with type-T thermocouples (TCs) at several locations: adjacent to the trim heaters, 15 cm above the cell, and at the cell bottom. The thermocouple at the cell bottom (TC2) is used for temperature control. Cell temperature regulation uses proportional-integral-derivative (PID) feedback implemented in LabView  \cite{noauthor_labview_nodate} to achieve better than 0.1 K stability. The LXe volume during normal operation within the cell is about 400 cc.

During filling the cell is first cooled to the chosen set point, typically around 163K. The absolute calibration of the control TC temperature is obtained from the observed onset of the gas to liquid phase change and from the vapor pressure observed at a capacitive manometer (P1), with an absolute accuracy of 5 torr. The Xe supply bottle is cooled with LN in order to reduce the bottle pressure (P2) to approximately 30 psi. Control of the bottle pressure is through PID feedback to an associated kapton bottle heater, and once the setpoint is reached the bottle is then opened and Xe gas passes through a hot zirconium alloy heated getter gas purifier \cite{noauthor_saes_nodate} and into the cell where it condenses. The LXe purity is not measured in the cell as this parameter is not critical for our purposes. Past experience at SLAC indicate that similar test cells in the setup achieve impurity levels corresponding to of order a few ppb oxygen after a single pass through the purifier \cite{albert_improved_2014}, as measured by the electron lifetime.
 
 A 33 Ohm DPAK 25W power resistor \cite{noauthor_specsheet_nodate} (about 0.7 mm on a side) was installed in the cell. The resistor is composed of a thick film resistive layer deposited on an aluminum oxide (Al$_{2}$O$_{3}$) substrate. The substrate is connected to an electrically isolated tinned copper heat sink that acts as the boiling surface. Figure \ref{fig:chamber_zoom} shows the resistor mounted to the power lines with two beryllium copper connectors at the terminals. From the side, three 2\,mm deep holes were drilled through the plastic (1\,mm thick) and into the copper top plate (1\,mm deep). A miniature type-T TC was installed in each hole, touching the copper and fixed with low outgassing cryogenic epoxy (Masterbond EP29LPSP). PTFE cable insulation was used to cover the extending wires. During every measurement the temperature of the resistor, and the temperature and pressure of the LXe cell (P1 $\And$ TC2) are recorded at 1 Hz, including several minutes before and after the power is turned on and off. The TCs within the resistor are calibrated relative to the control TC of the LXe cell, resulting in a systematic uncertainty of $\pm~0.4$ K.
 
 The test cell for these experiments is shown in Figure~\ref{fig:chamber_zoom}. The power resistor heat sink is positioned facing upwards and is clearly visible through the two opposing mini-CF sapphire window viewports. The production of bubbles could be seen, and video recordings were made with a digital camera through the viewports. The cell setpoint temperature (T$_{bulk}$) could be adjusted as needed, within a range from above the Xe freezing point at 161 K to as high as 175 K. Independent vapor pressure adjustment was less convenient but was possible by adjustment of the Xe supply bottle pressure setpoint. Experiments were typically completed in less than 8 hours, after which the Xe was cryogenically recovered into the storage bottle.

A single measurement consists of the following sequence. A set voltage is applied to the resistor for a period of at least one minute. During this period we visually inspect and record on video whether nucleate boiling occurs at the surface of the resistor. The voltage is then turned down and we wait for at least another two minutes before starting the next measurement at a higher voltage. Two sets of measurements were taken at different (constant) pressures while keeping the temperature of the LXe cell volume stable.
The resistor and LXe cell return to a steady thermal state within about 20 seconds after turning on or off the voltage. This is observed both by visual inspection of the fluid state and from the recorded time history of the TCs within the resistor.

\section{Simulations}
\begin{figure}[ht]
	\centering
	\captionsetup{}
	\includegraphics[width=0.32\textwidth]{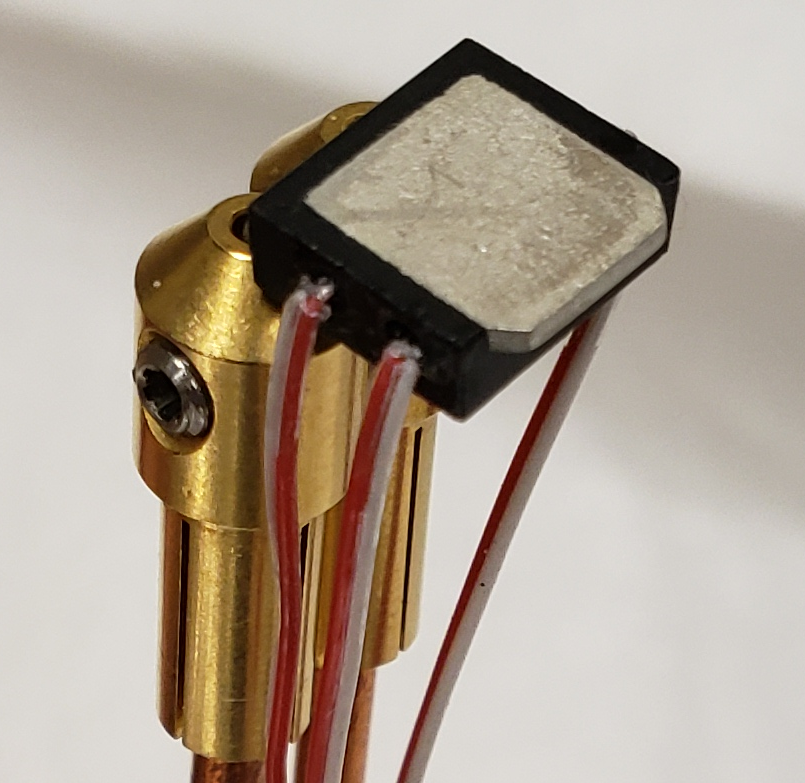}
	\includegraphics[width=0.32\textwidth]{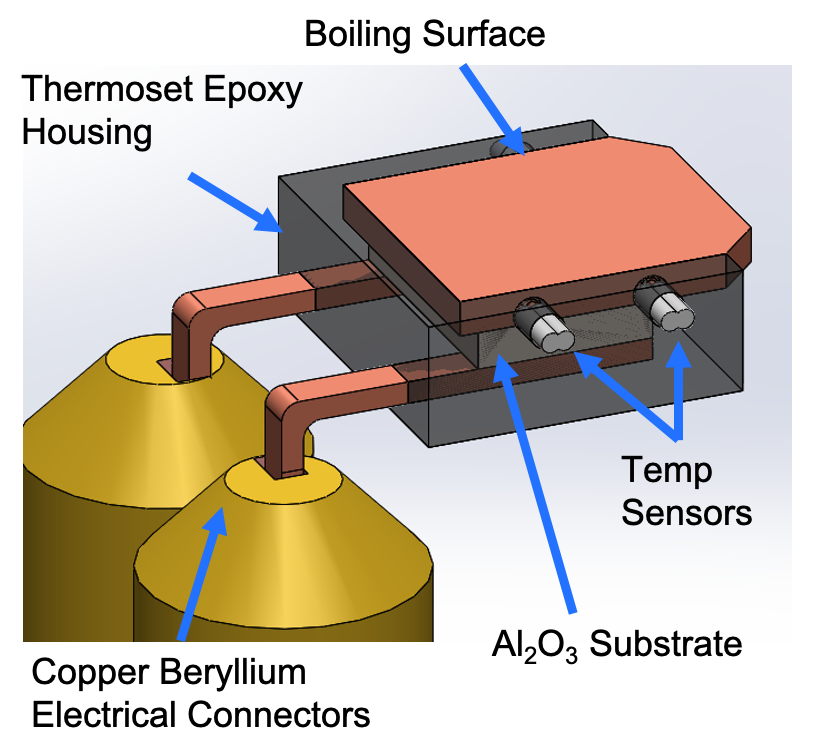}
	\includegraphics[width=0.34\textwidth]{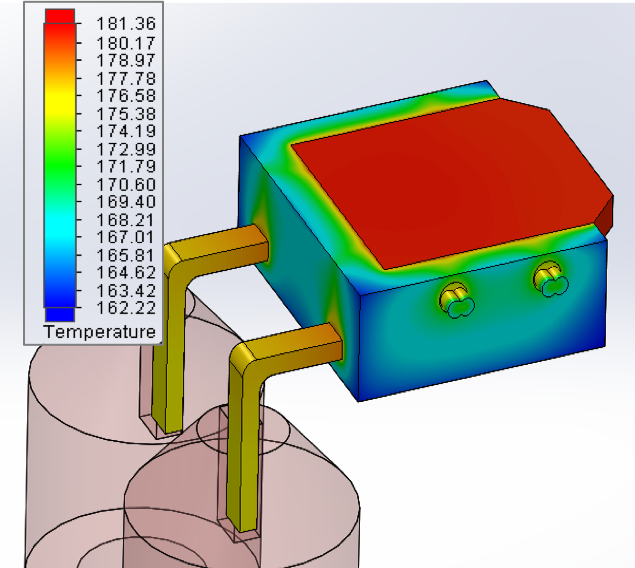}
	\caption{Left) Photo of the resistor mounted on beryllium copper connectors with the thermocouple wires.  Middle) CAD model showing resistor used for simulations. A thick film resistive layer is applied to the underside of the aluminum oxide substrate. Right) Plot of the CFD simulated surface temperature for an input power of 1.1 W and a pressure of 0.98 bar.}
	\label{fig:Resistor CAD}
\end{figure}

A dedicated CFD model was developed using SOLIDWORKS Flow Simulation (SW) \cite{noauthor_solidworks_2017}. The model determines the relationship between surface temperature and heat flux in the natural convection regime of LXe, and also investigates the temperature offset between the internal TCs and surface temperature of the resistor. The simulation model, shown in figure \ref{fig:Resistor CAD} is based on engineering drawings and material information from the resistor manufacturer Bourns \cite{noauthor_drawing_nodate} \cite{noauthor_materials_nodate} and includes the electrical connectors used in the experimental setup. The simulation does not include either the inner copper or outer stainless steel cryostat vessels, instead using a control volume of LXe roughly the size of the inner vessel with a liquid level 3 cm above the boiling surface.

The simulation is set up as a steady state external flow model with a stagnant starting fluid to imitate the pool boiling conditions of the experimental setup. The thick film resistor was modeled as a surface heat source placed directly on the alumina oxide substrate with a variable power output. The heat sink surface temperature, heat flux and global heat transfer rate were tracked and used to assess convergence \cite{sobachkin_numerical_2014}.

A thermal contact resistance was set for the connection between the resistor's electrical leads and the beryllium-copper connectors ($5 \times 10^{-4}$ m$^2$K/W), based on the SW recommended contact resistance ($R_t$) values for a contact pressure of 100 kN/m$^2$ \cite{noauthor_2017_nodate}. The heat flux and surface temperature of the top surface of the resistor were evaluated over the full range of applied power, up to the onset of boiling, and for both pressure settings.

Material selection for the resistor was based on information received from the manufacturer \cite{noauthor_materials_nodate}, and utilized predefined materials from the SW engineering database, with the exception of the thermoset epoxy (k = 1.7 W/mK) used for the resistor housing which was defined as a custom material. The LXe physical properties, obtained from the NIST database \cite{noauthor_xenon_nodate}, included density, thermal conductivity, specific heat, and viscosity.  

Additional parametric simulation studies were run to evaluate the effect of varying the TC location in the potted holes, the thermal conductivity of the resistor housing, and the thermal contact resistance between the resistor and the beryllium copper connectors. For the TC location, the proximity to the heating surface in the model was varied up to a maximum offset of 0.75 mm to look at the effect on TC temperature. The thermal conductivity value received for the housing is not temperature dependent like all the other materials in the model, and the contact resistance on the connectors is based on an unknown contact pressure. The selected range of values were informed by manufacturer data \cite{noauthor_materials_nodate} for thermal conductivity (1.2 to 2.1 W/mK) and the SW knowledge base \cite{noauthor_2017_nodate} for contact resistance ($2.6 \times 10^{-4}$ to $20.0 \times 10^{-4}$ m$^2$K/W). The results of these parametric studies are shown as the uncertainty bands from simulation in Figure~\ref{fig:temp_power}. 

\section{Results}
\begin{figure}[ht]
	\centering
	\captionsetup{}
	\includegraphics[width=\textwidth]{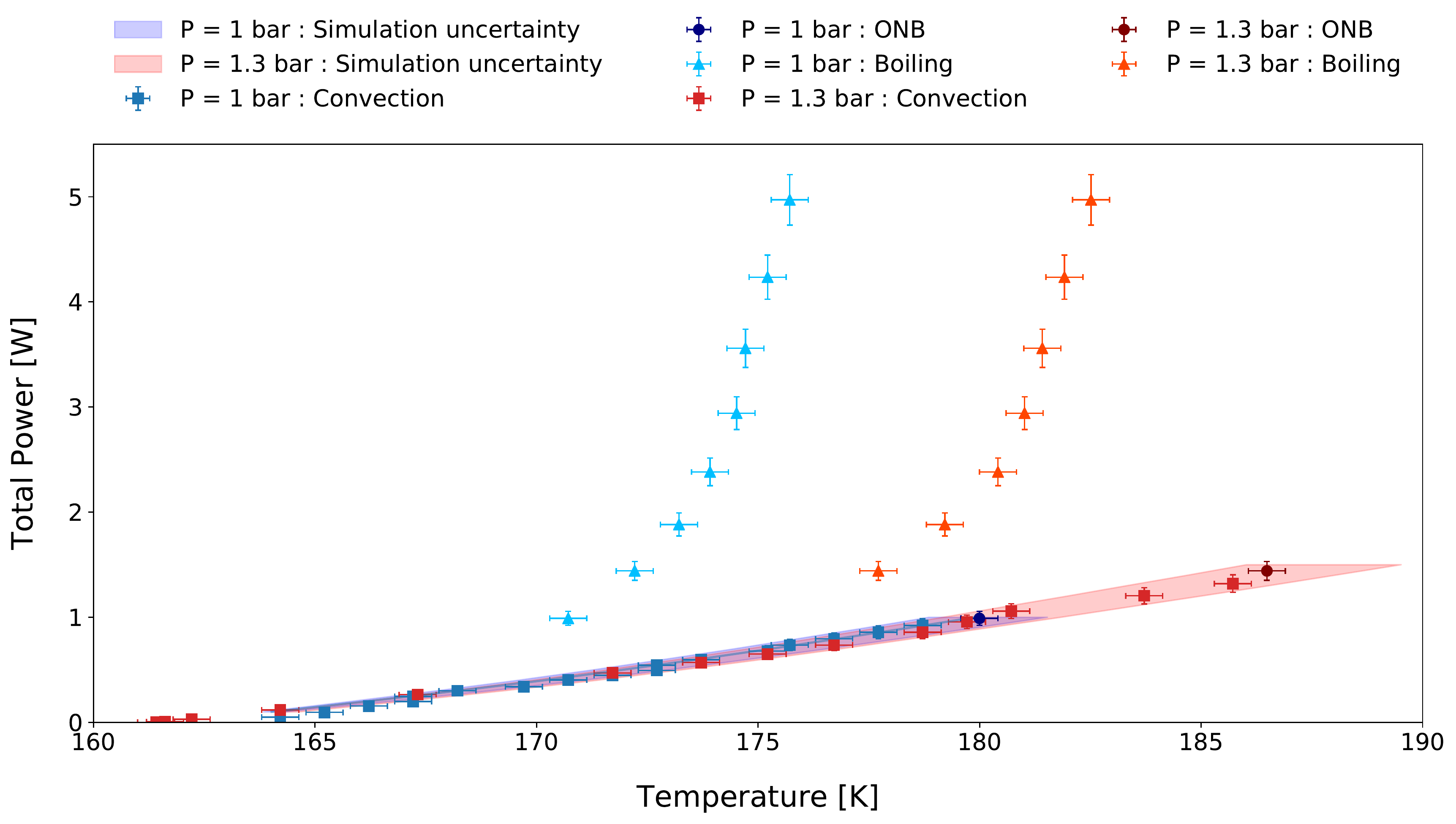}
	\caption{Measured temperature within the resistor for 1 bar (blue data points) and 1.3 bar (red data points) for a total power between 0 and 5 W with simulation results (colored bands) shown for the natural convection region. For increasing power the temperature within the resistor increases up to the point where nucleate boiling (triangular markers) takes over from convection (square markers) as the dominant method of heat transfer. Boiling is a more efficient form of heat transfer, causing the temperature within the resistor to sharply decrease before increasing again. The results from the simulation model examining the uncertainty on the temperature-power relation are shown by the colored bands for both pressures.}
	\label{fig:temp_power}
\end{figure}

Figure \ref{fig:temp_power} shows the total power through the resistor in LXe as a function of the TC temperature (T$_{tc}$) within the resistor for two different measurement sets: 1st measurement series with P = (0.98$\pm$0.02) bar and T$_{bulk}$ = (161.8$\pm$0.5) K, and 2nd measurement series at P = (1.32$^{+0.05}_{-0.01}$) bar and T$_{bulk}$ = (162.2$\pm$0.6) K. 
At $\sim$1 bar (all blue data points) the power was incrementally increased from 0 W up to about 5 W. The temperature of the resistor increases up to T$_{tc} = (178.7 \pm 0.4) $\,K without showing any signs of nucleate boiling and only convection within the LXe is observed (square markers). Nucleate boiling is first observed at $(0.99 \pm 0.06)$\,W and T=$(180.0\pm0.4)$\,K (round marker). Because nucleate boiling (triangular markers) is more efficient at heat transport than convection, the temperature within the resistor drops down to T$_{tc} = (170.7 \pm 0.4) $\,K before increasing again as the voltage increases. 
The data taken at 1.3 bar of pressure (all red data points) shows similar behavior. For this elevated pressure, convection (square markers) remains the only method of heat transport up to a temperature of T$_{tc} = 186.5 \pm 0.4 $\,K and a power of $1.32 \pm 0.06$\,W. For all higher power measurements nucleate boiling is observed (triangular markers).

The current going through the resistor is converted into heat in the resistive thick film, but only a certain fraction of this heat escapes through the top copper plate. To determine the heat flux (in W/cm$^2$) at the top surface of the resistor, where nucleate boiling occurs, the fluid simulation results are used. The simulation shows that at 1 bar and 1 W, 34.2\% of the heats flows through the top plate, while 30.1\% is dissipated through the plastic body and 35.7\% through the terminals on the back. The total power versus measured temperature data can now be converted into the average surface heat flux versus superheated wall temperature in LXe.

\begin{figure}[ht]
	\centering
	\captionsetup{}
	\includegraphics[width=\textwidth]{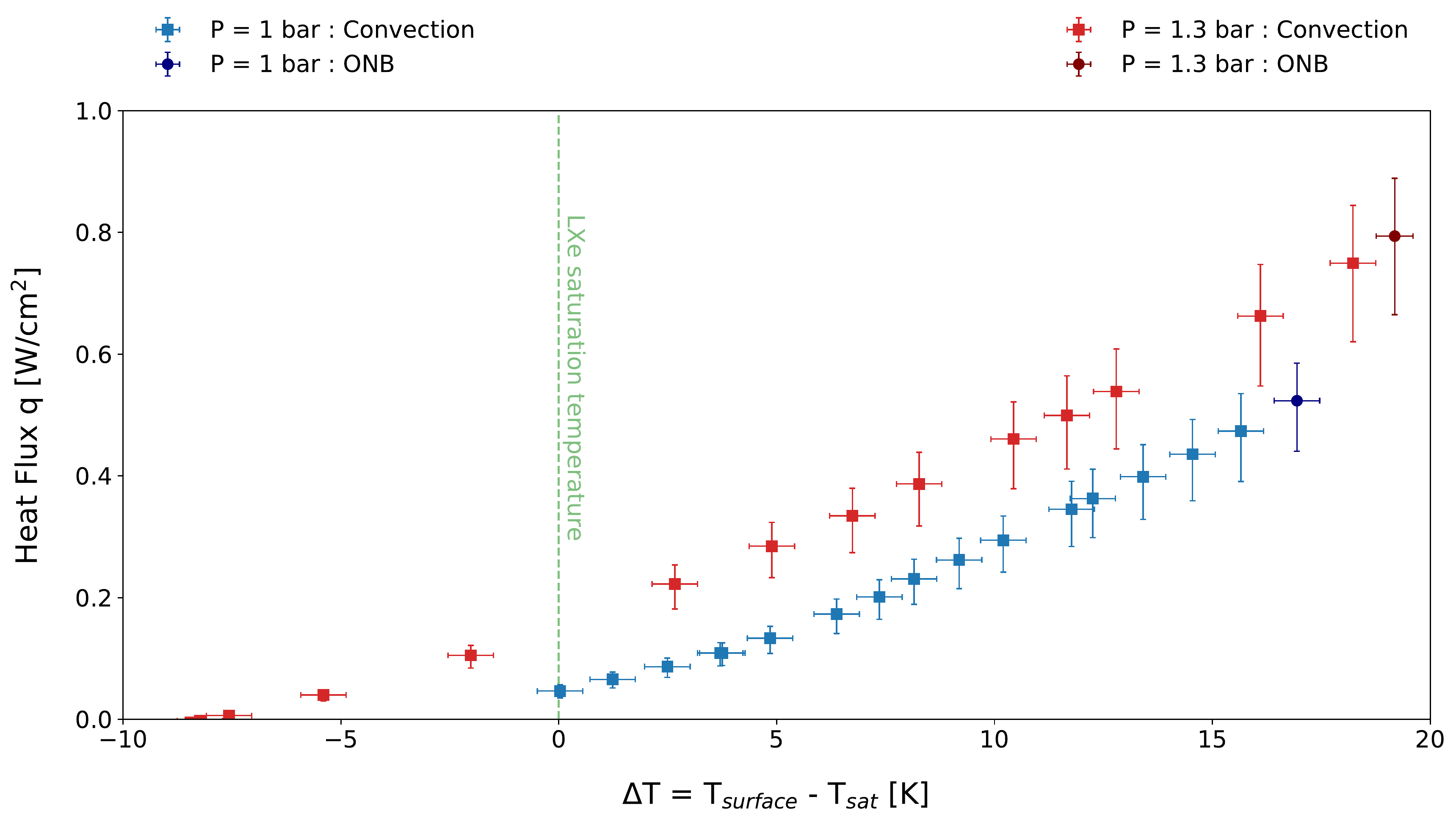}
	\caption{Measured data from figure \ref{fig:temp_power} combined with the results from the simulation model to show the relationship between the heat flux and $\Delta T$ for the convection region where no bubbles appear. The vertical dashed green line depicts the LXe saturation temperature. At both pressures the surface of the resistor can be heated well above the saturation temperature. The individual error bars on the data points include both the measurement uncertainties and the uncertainties from the simulation model.}
	\label{fig:deltat_q}
\end{figure}

Figure \ref{fig:deltat_q} shows the heat flux through the top surface of the resistor versus $\Delta$T for natural convection and the ONB at 1 bar (blue markers) and 1.3 bar (red markers). The total power through the resistor has been converted into the average heat flux on the top surface determined by the CFD model. The possible systematic bias from the uncertainty of both the thermal conductivity of the resistor housing and the thermal resistance of the terminals are added in quadrature to obtain the total uncertainty in the heat flux, represented by the y-axis error bars. The measurements show cooling by convection for superheat temperatures of $\Delta$T $<~16$ K, and $\Delta$T $<~18$ K, for 1 and 1.3 bar respectively. Above these temperatures boiling occurs. The exact values found for the ONB for the two different pressure measurements are shown in table \ref{tab:result}, together with their uncertainties from both the systematical and the statistical components. 

\begin{table*}[htb]
\centering
\begin{tabular}{c|c|c|c|c|c}
P [Bar] & T$_{bulk}$ [K] & $\Delta T_{wall,ONB}$ [K] & q [W/cm$^2$] & T$_{surface}$ [K] & T$_{sat}$ [K]\\
\hline
0.98$\pm$0.02 & 161.8$\pm$0.5 & 16.9$\pm$0.5 & 0.52$^{+0.06}_{-0.08}$ & 181.8$\pm$0.4 & 164.9$\pm$0.3\\
1.32$^{+0.05}_{-0.01}$ & 162.2$\pm$0.6 & 19.2$^{+0.4}_{-1.1}$ & 0.79$^{+0.09}_{-0.13}$ & 189.0$\pm{0.4}$ & 169.8$^{+1.0}_{-0.2}$\\
\hline
\end{tabular}
\caption{Measured values for the onset of boiling, as shown in figure \ref{fig:deltat_q}. The stated uncertainties include both the systematic and the statistical parts from readout and simulation. For the elevated pressure measurements a bias is included on the total pressure for a possible increase in the hydrostatic component due to the liquid level slowly rising while filling.}
\label{tab:result}
\end{table*}

\section{Discussion}
The measured value for $\Delta T_{wall,ONB}$ of (16.9$\pm$0.5) K at atmospheric pressure is within the range found by Haruyama \cite{haruyama_boiling_2002} of approximately 3.8 – 18 K. This, to our knowledge, is the only other published result for LXe. They do not quote uncertainties on their data points and it is therefore not possible to quantify the agreement between the two measurements. The heat flux at the ONB at atmospheric pressure we find is q = (0.79$^{+0.09}_{-0.13}$)~W/cm$^2$ which also fits within the approximate range of 0.4 – 2.5 W/cm$^2$ found by Haruyama. Their wide range could be explained by the fact that they include hysteresis and do not seem to account for any heat leaks in their system. Because experiments looking for rare event searches in LXe want to remain in the convection region, we have only focused on the onset of boiling when increasing the heat flux and excluded hysteresis effects when decreasing the heat flux. Our detailed modeling of the resistor components within the LXe test cell have shown that heat leaks are an important factor when determining the heat flux and should thus be taken into account.

The measured superheat values for LXe are much higher than those reported for other fluids like water by Lamarsh of 5 K \cite{lamarsh_introduction_1983}, and of nitrogen by Bouazaoui et al.~of 11.3 K \cite{bouazaoui_experimental_2017}. As the onset of boiling is dependent on the fluid properties (such as the bubble angle, density, viscosity, etc.) this is not unexpected. Unfortunately, as of yet, no theoretical model exists that can predict the ONB solely from first principles and fluid properties.

As there are clear examples showing that the boiling heat transfer curve is dependent on both the pressure \cite{kosky_pool_1968} and the amount of subcooling \cite{warrier_onset_2002} (temperature of bulk fluid below saturation temperature), we present the values separately in table \ref{tab:result}. Further experiments are required to better determine the effect of changing the pressure, the amount of subcooling and the surface roughness on the onset of boiling in LXe.

\section{Conclusion and Outlook}
The results of this experiment show that the ONB in LXe occurs at surface temperatures of (16.9$\pm$0.5) K and (19.2$^{+0.4}_{-1.1}$) K at pressures of P=$(0.98\pm0.02)$\,bar and P=$(1.32^{+0.05}_{-0.01})$\,bar, respectively.
These results were obtained using a horizontal flat copper resistor surface inside a LXe cell cooled to 162 K, with a detailed CFD model used to accurately estimate the surface temperature and heat flux. This is the first result showing the ONB parameters in LXe with an uncertainty of a few degrees, including both subcooling and increased pressure.

These measurements of the ONB parameters for LXe can now be applied to different experimental designs. In order to evaluate if a specific component is likely to initiate boiling it is necessary to estimate the surface temperature in LXe and compare the difference from saturation temperature to the superheat at ONB ($\Delta$T$_{wall,ONB}$). The estimation of the surface temperature can be done through CFD simulations of the geometry, materials, and heat flux. Similarly, the combination of CFD simulations and the knowledge of $\Delta$T$_{wall,ONB}$ can be used to design detector components such that nucleate boiling is avoided.

\section*{Acknowledgements}
We are grateful for the useful discussions with, and ongoing support of John Orrell, Knut Skarpaas VIII, Bob Conley and our many collaborators. Initial CFD modeling identifying the potential for boiling was supported by Laboratory Directed Research and Development (LDRD) through the Nuclear-physics, Particle-physics, Astrophysics, and Cosmology (NPAC) Initiative at Pacific Northwest National Laboratory (PNNL). PNNL is a multiprogram national laboratory operated by Battelle for the U.S. Department of Energy under Contract No. DE-AC05-76RL01830. Subsequent modeling and analysis of the experimental setup presented here was supported through a grant from the U.S. Department of Energy Office of Science, Office of Nuclear Physics to support ton-scale neutrinoless double beta decay experimental Research and Development (R\&D). SLAC National Accelerator Laboratory, is supported by the U.S. Department of Energy, Office of Science under Contract No. DE-AC02-76SF00515.



\bibliographystyle{unsrturl}
\bibliography{BoilingReferences.bib}

\end{document}